\newcommand{\cM}{{\sl M}}
\newcommand{\mass}{{\cal M}}
\newcommand{\tM}{{\widetilde M}}

\newcommand{\cL}{{\cal L}}
\newcommand{\cw}{{ w}}

\newcommand{\be}{\begin{equation}}
\newcommand{\ee}{\end{equation}}
\newcommand{\ba}{\begin{eqnarray}}
\newcommand{\ea}{\end{eqnarray}}
\newcommand{\non}{\nonumber\\ }

\documentstyle[12pt,fullpage]{article}
\begin{document}

\renewcommand{\thefootnote}{\fnsymbol{footnote}}
\font\csc=cmcsc10 scaled\magstep1
{\baselineskip=14pt
 \rightline{
 \vbox{\hbox{YITP-97-19}
       \hbox{May 1997}
}}}

\vfill
\begin{center}
{\large\bf
Lorentz Symmetry of Supermembrane\\
in Light Cone Gauge Formulation
}

\vfill

{\csc Kiyoshi EZAWA}\footnote{JSPS fellow}\setcounter{footnote}
{0}\renewcommand{\thefootnote}{\arabic{footnote}}\footnote{
      e-mail address : ezawa@yukawa.kyoto-u.ac.jp},
{\csc Yutaka MATSUO}\footnote{
      e-mail address : yutaka@yukawa.kyoto-u.ac.jp},
{\csc Koichi MURAKAMI}\footnote{
      e-mail address : murakami@yukawa.kyoto-u.ac.jp}\\
\vskip.1in

{\baselineskip=15pt
\vskip.1in
  Yukawa Institute for Theoretical Physics \\
  Kyoto University, Sakyo-ku, Kyoto 606-01, Japan \\
\vskip.1in
}

\end{center}
\vfill

\begin{abstract}
{
We prove the Lorentz symmetry
of supermembrane theory
in the light cone gauge to complete the program
initiated by de Wit, Marquard and Nicolai.
We give some comments on extending the formulation
to the M(atrix) theory.
}
\end{abstract}
\vfill

hep-th/9705005
\setcounter{footnote}{0}
\renewcommand{\thefootnote}{\arabic{footnote}}
\newpage
\vfill

\section{Introduction}


After the discovery of the string duality, 
our perception of the string theory was drastically 
changed.  What used to be the obscure inhabitants of the string theory,
the $p$-branes, turned out to be the key ingredients of
the non-perturbative physics.
$\cM$-theory is supposed to be one of the most symmetric form of the
``string'' theory.  However, because of our ignorance of the quantization
of the $p$-branes, the very definition of the theory
has been largely unknown.

By critical use of the simplification 
due to the infinite momentum frame, 
BFSS \cite{r:BFSS} proposed a constructive
definition of the $\cM$-theory.  The momentum
along the eleventh dimension is identified with the zero-brane charge.
The infinite boost kills the degree of freedom which has
zero (fundamental string) and negative (anti-zero brane) charges.
The resulting Lagrangian is made up only with the
zero-branes described by the large $N$ limit of the $SU(N)$
Yang-Mills theory.
BFSS have indicated two major evidences
which support their idea.
\begin{enumerate}
\item The matrix theory Lagrangian coincides with that of
supermembrane proposed by de Wit, Hoppe and Nicolai (dWHN)
\cite{r:dWHN}
if one replaces the gauge group from $SU(N)$ to the 
area preserving diffeomorphism (APD) in two dimensions.
\item The scattering of the zero-branes coincides with
the prediction of the eleven dimensional supergravity.
\end{enumerate}
%

As usual, the subtlety in the infinite momentum frame is
the Lorentz symmetry.
This problem is very difficult to analyze in the matrix theory 
since the momentum exchange in the eleventh dimension
means the exchange of zero-brane charge.
We need to treat the quantum process which changes the
size of matrices\footnote{
Beautiful treatment of this issue is recently 
proposed by Polchinski and Pouliot \cite{r:PP} by considering
scattering of two membranes where zero-brane charge
can be treated as the monopole charge on the world brane.
In this setting, the zero-brane exchange can
be calculated by the instanton calculus on the world brane.}.  

On the other hand, the analysis of the similar problem
in the dWHN model is accessible since we know
the covariant Lagrangian in eleven dimensions.
Indeed this program was nearly accomplished by
de Wit, Marquard and Nicolai (dWMN)\cite{r:dWMN}.
They defined the Lorentz generators
and have shown that they commute with
the Hamiltonian of the system.
In their proof, they essentially used
various identities of the APD 
tensors.  

The purpose of this technical note is to complete this program,
namely to give the direct computation of 
the algebra of the Lorentz generators.
Our result supports the Lorentz symmetry
after the cancellation among numerous non-trivial factors.
We need to prove some additional identities of the APD
tensors to finalize our result.
In section two, we briefly review the result of
dWMN to make this note self-contained.
In section three, we summarize
our proof of the Lorentz invariance.
In section four, we give a discussion on 
the possible extension of our result
to the M(atrix) theory.  
One of the generators of Lorentz algebra depends essentially
on the metric of the membrane world volume.
Therefore, it is not invariant under the APD and it
causes some nontriviality. We 
argue that this fact might give 
a hint to eleven dimensional
definition of \cM-theory.
Explicit computations
and technical comments are provided in the appendix.
In appendix A, we describe the identities
between the APD tensors.
In appendix B, we summarize properties of
the Clifford algebra of $SO(9)$.
The identities in these sections are used in 
appendices C and D to prove the Lorentz algebra.

\section{Summary of dWHN model}
DWHN model \cite{r:dWHN} is defined as a $0+1$ dimensional 
supersymmetric Yang-Mills system whose gauge group is
the APD of a fixed two dimensional manifold.
The Lagrangian (slightly modified from
the original definition) is
\be
\sqrt{\cw}^{-1}\cL = \frac{1}{2}(D_0 {\vec X})^2
+\frac{i}{2}\theta D_0 \theta -\frac{1}{4}
\left(\left\{X^a, X^b\right\}\right)^2
+\frac{i}{2}\theta\gamma_a\left\{X^a,\theta\right\},
\ee
where the definitions of the notation are following.
$X^a(t,\sigma^r)$, $\theta_\alpha(t,\sigma^r)$ 
($a=1,\ldots,9$, $\alpha=1,\ldots,16$, $r=1,2$)
are the quantum mechanical variables whose internal 
degree of freedom is described by two parameters $\sigma$.
The indices $a$ and $\alpha$ are respectively
the vector and the spinor degrees of freedom of $SO(9)$.
$\cw_{ij}$ is the $2\times 2$ 
metric tensor for the parameter space
and $\cw$ is its determinant.
The curly bracket,
$
\left\{A,B\right\}
\equiv \frac{\epsilon^{rs}}{\sqrt{\cw(\sigma)}}
\partial_r A(\sigma) \partial_s B(\sigma),
$
and the covariant derivative,
$
D_0 X^a=\partial_0 X^a-\left\{\omega,X^a\right\}
$,
$
D_0 \theta=\partial_0 \theta-\left\{\omega,\theta\right\},
$
define the gauge transformation based on the APD,
$$
\delta X^a=\left\{\xi, X^a\right\},\quad
\delta \theta=\left\{\xi, \theta\right\},\quad
\delta\omega=\partial_0\xi+\left\{\xi, \omega\right\}.
$$

The canonical Hamiltonian  is
\ba
H &=& -\int d^2 \sigma P^{-}(\sigma)\non
&=& \frac{1}{P^+_0}
\int d^2 \sigma
\sqrt{\cw(\sigma)}
\left(
\frac{1}{2}\cw^{-1}\vec{P}^2+\frac{1}{4}
\left(\left\{X^a,X^b\right\}\right)^2
{} -\frac{i}{2}\theta\gamma_a\left\{X^a,\theta\right\}
\right).
\ea
$\vec{P}$ denotes the canonical momentum conjugate to
$\vec{X}$.
The non-vanishing Dirac brackets are
\ba
\left(X^a(\sigma),P^b(\rho)\right)_{DB}
& = & \delta^{ab}\delta^{(2)}(\sigma,\rho),\non
\left(\theta_\alpha(\sigma),\theta_\beta(\rho)\right)_{DB}
&=& -\frac{i}{\sqrt{\cw(\sigma)}}\delta_{\alpha\beta}
\delta^{(2)}(\sigma,\rho).
\ea

The Gauss law constraints associated with the APD
can be written as $\varphi(\sigma) \approx 0$
and $\varphi^\lambda \approx 0$ 
where
\ba
\varphi(\sigma) &\equiv&
{} - \left\{ \frac{P^a(\sigma)}{\cw(\sigma)},
{X^a}(\sigma)\right\}
{} - \frac{i}{2}\left\{
\theta(\sigma),\theta(\sigma)\right\},\non
\varphi^\lambda & \equiv &
{} -\int d^2 \sigma \Phi^{(\lambda)r}(\sigma)
\left(\vec{P}(\sigma)\cdot\partial_r \vec{X}(\sigma)
+\frac{i}{2}\sqrt{\cw(\sigma)}\theta(\sigma)\partial_r
\theta(\sigma)\right).
\ea
$\Phi^{(\lambda)r}$ is the basis of
the harmonic vectors in the parameter space\cite{r:dWMN}.

The light cone directions are expressed through
$X^\pm=\frac{1}{\sqrt{2}}(X^{10}\pm X^0)$ where
one of them is identified with the world volume
time variable
$X^+(\tau)= X^+(0)+\tau$.
The other one is defined by
\be
\partial_r X^-(\sigma) 
=
{} -\frac{1}{P^+_0}\left(
\frac{1}{\sqrt{\cw(\sigma)}}
\vec{P}(\sigma)\cdot\partial_r\vec{X}(\sigma)
+\frac{i}{2}\theta(\sigma)\partial_r\theta(\sigma)\right).
\ee
The integrability conditions of this differential equation
coincide with the Gauss law constraints. When integrated, it gives
\be
X^{-}(\sigma) =
q^- - \frac{1}{P^+_0}\int d^2 \rho G^{r}(\sigma, \rho)
\left( \vec{P}(\rho)\cdot \partial_r\vec{X}(\rho) +
\frac{i}{2}\sqrt{\cw(\rho)} \theta(\rho)\partial_r \theta(\rho)
\right),
\ee
where the integration constant
satisfies $(q^-,P^+_0)_{DB}=1$ and
$G^r(\sigma, \rho)$ is the Green function
defined by
$D^{\rho}_rG^r(\sigma,\rho) = -(\cw(\sigma))^{-1/2} 
\delta^{(2)}(\sigma, \rho) +1
$.

This system has supersymmetry generated by
\ba
Q^+ & = & \frac{1}{\sqrt{P^+_0}}\int d^2\sigma\left(
P^a\gamma_a+ \frac{\sqrt{\cw}}{2} \left\{
X^a, X^b\right\} \gamma_{ab} \right)
\theta,\non
Q^- & = & \sqrt{P_0^+}\int d^2\sigma \sqrt{\cw(\sigma)}
\theta.
\ea

The Lorentz generators are defined by,
\ba
M^{ab} & = & \int d^2\sigma
\left( -P^aX^b + P^b X^a - \frac{i}{4} \theta \gamma^{ab}
\theta\right),\non
M^{+-} & = & \int d^2 \sigma \left(
{} -P^+X^- + P^- X^+\right),\non
M^{+a} & = & \int d^2 \sigma (-P^+ X^a + P^a X^+),
\\
M^{-a} & = & \int d^2\sigma \left(
P^a X^- - P^-X^a - \frac{i}{4P^+_0}
\theta\gamma^{ab}\theta P_b
{} - \frac{i\sqrt{\cw}}{8P^+_0} 
\left\{ X_b, X_c\right\} \theta \gamma^{abc}
\theta \right).\nonumber
\ea

DWMN \cite{r:dWMN}  proved that
these generators satisfy,
\be
\frac{d}{d\tau} M = 
\frac{\partial}{\partial \tau} M +
(M, H)_{DB}=0,
\ee
namely the conservation of these charges.
Although this is a nontrivial consistency check, 
it is obviously  important to prove that the
Dirac brackets between these charges 
indeed satisfy the Lorentz algebra.
%

As in \cite{r:dWMN}, we carried out our computation by using 
mode expansion in two dimensional parameter space.
To define the basis, we pick the covariant Laplacian
in the parameter space and define the basis as
its eigenfunctions,
\be
\Delta Y_0 = 0, \quad
\Delta Y_A = -\omega_A Y_A
\ee
where $\omega_A >0$\footnote{We consider the
case when the parameter space is compact and the
spectrum is discrete.}.
The index A will take value in positive integers.
When we need to treat both zero- and non-zero modes
we will use indices I,J.
We may require them to satisfy the orthonormal condition,
\be
\int d^2 \sigma \sqrt{\cw(\sigma)}
Y^I(\sigma) Y_J(\sigma) = {\delta^I}_{J}
\qquad
Y^I\equiv Y_I^* = \eta^{IJ}Y_J.
\ee
The completeness condition is
\be
\sum_{A} Y^A(\sigma) Y_A(\rho) =
\frac{1}{\sqrt{\cw(\sigma)}}\delta^{(2)}(\sigma ,\rho)
{} -1.
\ee
All fields will be expanded in terms of $Y_I$ 
such as, $X^a(\sigma) = \sum_{I} X^{aI}Y_I(\sigma)$.
The Green function which appeared in the definition
of $X^-$ is then expanded as,
\be
G^r(\sigma, \rho) = \sum_A
\frac{1}{\omega_A}Y^A(\sigma) \partial^r Y_A(\rho).
\ee

The structure constant of the APD is given by
\be
f_{ABC} \equiv \int d^2\sigma 
\sqrt{\cw(\sigma)} Y_A(\sigma) \left\{
Y_B(\sigma),Y_A(\sigma) \right\},
\quad
\left\{ Y_A,Y_B\right\}
= f_{ABC}Y^C.\label{eq:fabc}
\ee
We also define,
\ba
d_{ABC} & = & \int d^2\sigma
\sqrt{\cw(\sigma)} Y_A(\sigma)Y_B(\sigma)Y_C(\sigma)\non
c_{ABC} & = & 
{} -2\int d^2\sigma
\sqrt{\cw(\sigma)} \frac{\cw^{rs}}{\omega_A}
\partial_rY_A Y_B \partial_s Y_C.\label{eq:dabc}
\ea
The tensor $c_{ABC}$ is motivated to express
the mode expansion of the Green's function
and is indispensable to express $X^-$. 
Whereas the tensors $f$ and $d$ are invariant under
APD, $c$ is not invariant because it depends explicitly
on the metric.
DWMN have argued that there is no
modification  which makes it invariant.
In this sense, it is a challenge to find the analogue of
this constant when we treat the M(atrix) theory.
We will come back to this issue later.

In terms of the coefficients of mode expansion,
the Dirac brackets are rewritten as,
\ba
(X^a_A, P^b_B)_{DB} & = & \delta^{ab}\eta_{AB},\non
(\theta_{\alpha A}, \theta_{\beta B})_{DB}
& = & -i\delta_{\alpha,\beta} \eta_{AB},\non
(q^-,P^+_0)_{DB} & = & 1,\non
(X^a_0, P^b_0)_{DB} & = & \delta^{ab} , \non
(\theta_{\alpha 0},\theta_{\beta 0})_{DB}
& = & -i \delta_{\alpha,\beta}.
\label{eq:DB}
\ea

Let us write down the mode expansion of various
conserved charges. 
The elements of the APD
are given by,
\ba
\varphi_A & = & f_{ABC}\left(
\vec{X}^B\cdot\vec{P}^C -\frac{i}{2}\theta^B\theta^C\right)
\non
\varphi_\lambda & = & f_{\lambda BC}
\left(
\vec{X}^B\cdot\vec{P}^C -\frac{i}{2}\theta^B\theta^C\right)
\ea
As dWMN indicated, to describe Lorentz generators,
it is convenient to separate the zero-mode and others.
\ba
H & = & \frac{\vec{P}_0^2}{2P^+_0}+\frac{
\mass^2}{2P^+_0} \non
\mass^2 & = & \vec{P}_A^2+\frac{1}{2} (
f_{ABC}X^B_aX^C_b)^2-
if_{ABC} \theta^A \gamma^aX^B_a\theta^C
\label{eq:M2}
\\
M^{ab} & = & -P^a_0X^b_0+P^b_0X^a_0
{} -\frac{i}{4} \theta_0\gamma^{ab} \theta_0
{} -P^a_AX^{bA}+P^b_AX^{aA} -\frac{i}{4}\theta_A
\gamma^{ab}\theta^A\non
M^{+-} & = & - P^+_0q^- -H\tau,\non
M^{+a} & = & -P_0^+X^a_0+\tau P^a_0.
\ea
We write $M^{-a}$ in the following form,
\be
M^{-a} = (M^{-a})^{(0)} 
+\frac{1}{P_0^+}\left(
P_{0b} \tM^{ab}-\frac{i}{2} \theta_0 \gamma^a
\widetilde{Q}^+\right)
+\frac{1}{P^+_0}\tM^{-a},
\ee
where
\ba
(M^{-a})^{(0)} & = & q^- P^a_0+X^a_0H -\frac{i}{4 P^+_0}
\theta_0\gamma^{ab}\theta_0P^b_0,\non
\widetilde{Q}^+ & = & 
(P^a_A+\frac{1}{2}f_{ABC}X^B_aX^C_b\gamma^{ab}
)\theta^A,\non
\tM^{ab} & = & -P^a_AX^{bA}
+ P^b_AX^{aA} -\frac{i}{4}\theta_A\gamma^{ab}\theta^A\non
\tM^{-a} & = & \frac{1}{2} d^{ABC}X^a_A(
\vec{P}_B\cdot \vec{P}_C + \frac{1}{2} 
(f_B^{DE}X^b_DX^c_E)(f_C^{FG}X^b_FX^c_G)
{} -i f_C^{DE}X^b_D\theta_B\gamma_b\theta_E)\non
&&-\frac{i}{4}d^{ABC}P_{Ab}\theta_B\gamma^{ab}\theta_C
+\frac{1}{2}c^{ABC} P^a_A(\vec{P}_B\cdot
\vec{X}_C+\frac{i}{2}\theta_B\theta_C)\non
&& -\frac{i}{8}f^{ABC} d_A^{DE}X_{Bb}
X_{Cc}\theta_D\gamma^{abc}\theta_E.
\label{eq:Mt}
\ea

\section{Lorentz symmetry in supermembrane}

In this section we summarize our proof of the
Lorentz symmetry. The detail is explained in appendices C and D.
What we want to do is to show that we have the eleven dimensional
Lorentz algebra:
\begin{equation}
(M^{\mu\nu},M^{\rho\sigma})_{DB}
=\eta^{\mu\rho}M^{\nu\sigma}+\eta^{\nu\sigma}M^{\mu\rho}
     -\eta^{\mu\sigma}M^{\nu\rho}-\eta^{\nu\rho}M^{\mu\sigma},
  \label{eq:11dim}
\end{equation}
where the indices $\mu, \nu, \rho$ and $\sigma$ run the eleven
dimensional space-time indices $+,-,1,\cdots ,9$.

DWMN  have shown
\begin{eqnarray}
(\tilde{M}^{ab},{\cal M}^{2})_{DB}&=&0,\\
(\tilde{Q}^{+},{\cal M}^{2})_{DB}&=&2\theta_A\varphi^A,\\
(\tilde{M}^{-a},{\cal M}^{2})_{DB}&=&-
\left({f_B}^{DE}X^a_D\vec{X}_C\cdot\vec{X}_E
{} -\frac{i}{2}\theta_B\gamma^a\theta_C\right)
(c^{ABC}\varphi_A+c^{\lambda BC}\varphi_\lambda) .
\end{eqnarray}
The RHS vanishes in the physical subspace.
By using these relations and the
Dirac bracket $(q^{-},P^{0}_{+})_{DB} = 1$,
we can easily prove the Lorentz algebra(\ref{eq:11dim}) except
$(M^{-a},M^{-b})_{DB}=0$, modulo the first class constraints.

Proof of the only nontrivial part $(M^{-a},M^{-b})_{DB}=0$
goes as follows.
By separating zero and non-zero modes we find
\footnote{In this calculation we have used the
relation:
$$
\left( (M^{-a})^{(0)},(M^{-b})^{(0)}\right)_{DB}
  =\frac{i}{4(P^{+}_{0})^{2}}\theta_{0}\gamma^{ab}\theta_{0}
      {\cal M}^{2}.
$$}
\begin{equation}
\label{eq:--}
(M^{-a},M^{-b})_{DB}=\frac{1}{(P^+_{0})^{2}}(C+D),
\end{equation}
where
\begin{eqnarray}
C&=&(-{\cal M}^{2}\tilde{M}^{ab}
      -\frac{i}{4}\tilde{Q}^{+}\gamma^{ab}\tilde{Q}^{+})
     +(\tilde{M}^{-a},\tilde{M}^{-b})_{DB} \ , 
\label{eq:C}\\
D&=&-\frac{i}{2}\theta_{0}\gamma^{a}
     (\tilde{Q}^{+},\tilde{M}^{-b})_{DB}
     +\frac{i}{2}\theta_{0}\gamma^{b}(\tilde{Q}^{+},\tilde{M}^{-a})_{DB}.
\end{eqnarray}

In the appendices C and D
we will show that the following relations hold 
modulo the first class constraints,
\ba
\label{eq:MM}
C&=&0,\\
\label{eq:QM}
(\tilde{Q}^{+},\tilde{M}^{-a})_{DB}&=&0.
\ea
Here we only quote the final result.

\begin{enumerate}
\item The first equation (\ref{eq:MM}):

\ba
&& \frac{i}{2}\left(
X^a_A(\theta^A\gamma^b\theta_D) - X^b
(\theta^A\gamma^a\theta_D)\right)
\varphi^D\non
&&-\frac{i}{4} \left(
X^a_A(\theta_B\gamma^b\theta_E)-
X^b_A(\theta_B\gamma^a\theta_E)\right)
d^{ABC}({c_{DC}}^E\varphi^D+{c_{\lambda C}}^E\varphi^\lambda)\non
&&+\frac{i}{2}(\theta_D\gamma^{abd}\theta^D)X_{dE}\varphi^E\non
&&-\frac{i}{4}(\theta_D\gamma^{abd}\theta_E)X_{dC}
{d_A}^{DE}({c_B}^{AC}\varphi^B+{c_{\lambda}}^{AC}\varphi^\lambda)\non
&& +\left(
{} -\frac{1}{\omega_A}(\frac{1}{\omega_C}f^{CAB}\varphi_C
+f^{\lambda AB}\varphi_\lambda)+\frac{1}{2\omega_A\omega_B}
{f_C}^{AB}\varphi^C\right)
(P^a_AP^b_B-P^b_AP^a_B)\non
&& -\frac{1}{2} d^{AIC}{f_I}^{DE}(X^a_A X^b_E
{} -X^b_A X^a_E)\vec{X}_D\cdot\vec{X}_F
({c_{HC}}^F\varphi^H + {c_{\lambda C}}^F \varphi^\lambda)\non
&& -f^{DAE}(X^a_A X^b_E - X^b_A X^a_E)
\vec{X}_D\cdot \vec{X}_F\varphi^F.
\label{eq:MtMt}
\ea

\item The second equation (\ref{eq:QM}):

\be
\theta_{\alpha C}X^a_D {d^{BD}}_E\varphi^E
+\frac{1}{2} (\gamma^{ad}\theta_C)_\alpha
X_{dD}({c_E}^{CD}\varphi^E+{c_{\lambda}}^{CD}\varphi^\lambda).
\label{eq:QMt}
\ee
\end{enumerate}

In the physical subspace where
$\varphi^A,\varphi^\lambda\approx 0$ both of the equations vanish.
This completes the proof of the 
Lorentz invariance of the dWHN model.

\section{Discussion: Lorentz invariance of M(atrix) Theory}

Although our computation is rather tedious,
it has a merit that it can be carried out quite systematically.
Therefore we are eager to  speculate that 
such an analysis may be applicable
to prove the Lorentz invariance of the M(atrix) theory in the
large $N$ limit.

Indeed there are well known correspondence between
$SU(N)$ in the large $N$ limit and the APD.
{}For the simpler part, the translation table
is given as follows,
\vskip 5mm
\begin{center}
\begin{tabular}{|c||c|}\hline
APD & $SU(N)$  \\ \hline\hline
$Y^A(\sigma)$ & $T_A$ \\ \hline
$\left\{X,Y\right\}$ &  $\left[ X, Y\right]$\\ \hline
$\int d^2 \sigma \sqrt{\cw(\sigma)} 
X(\sigma)$ & $ \mbox{Tr }X$\\ \hline
$\left\{Y^A,Y^B\right\}={f^{AB}}_C Y^C$ & 
  $\left[ T^A, T^B\right] = {f^{AB}}_C T^C$\\ \hline
$Y^A Y^B = {d^{AB}}_C Y^C$ & $\left[T^A, T^B\right]_+ 
= {d^{AB}}_C T^C$
\\ \hline
\end{tabular}
\end{center}
\vskip 5mm

When we compute
the anti-commutator of $SO(9)$ supercharges to study
the appearance of $p$-branes of various dimensions,
these two theory were essentially the same
\cite{r:dWHN}\cite{r:BSS}
except for the vanishing of the five brane
charges in the supermembrane approach\footnote{
Some discrepancy observed in \cite{r:BSS} can be removed
when we carefully keep the Schwinger term in the matrix
computation. See appendix F for detail.}.
In such computation the corresponding generators
of the M(atrix) theory are available after the
use of above dictionary.
It means the computation does not depend on
particular geometry of the world volume.

{}For the calculation of the Lorentz invariance, on
the other hand, we need to introduce the third
tensor $c_{ABC}$ which depends explicitly on the metric
of the parameter space and the direct translation
becomes more involved.
Of course, when the geometry of 
the parameter space is fixed (say the Riemann
surface of genus $g$), we already know the
non-commutative analogue of the surface
which can be embedded in the large $N$ limit 
of $SU(N)$\cite{r:etc}.
In such a situation, the construction of the
corresponding Lorentz generators in
M(atrix) theory  becomes
possible.
{}For that purpose, it is convenient
to indicate an identity for
the tensor $c_{ABC}$ (see the appendix E for the proof),
\be
\label{eq:cd}
c_{ABC} = \left(\frac{\omega_B-\omega_C}{\omega_A} -1\right)
d_{ABC}.
\ee
This formula shows that once the notion of the
Laplacian is generalized to the M(atrix) theory
we can construct the tensor $c$.  Because we take
the orthogonal basis $Y^I$ as the eigenfunction
of the Laplacian, the basis of $SU(N)$ should
be also taken from eigenvectors that diagonalize
such an operator.  If the Laplacian thus defined
have definite ``classical'' limit in $N\rightarrow \infty$,
the relations among the tensors $f$, $d$, $c$ will
also hold in this limit. This means we will recover the
Lorentz invariance.
We have now the second type of correspondence.

\vskip 5mm
\begin{center}
\begin{tabular}{|c||c|}\hline
APD & $SU(N)$  \\ \hline\hline
$\Delta$ & $\widetilde\Delta$\\ \hline
$\omega_A$ & $\widetilde\omega_A$ \\ \hline
$\Delta Y_A = -\omega_A Y_A$ &
$\widetilde\Delta T_A = -\widetilde\omega_A T_A$\\ \hline
\end{tabular}
\end{center}
\vskip 5mm

Let us illustrate the idea in a concrete example, namely
the case of the (non-commutative) torus.
In the APD case, the base of the mode expansion
is nothing but the Fourier expansion,
\ba
Y_A &=& \exp(i(A_1 \sigma_1+A_2\sigma_2)) ,\non
\Delta Y_A & = &  -\omega_A Y_A, \quad
\omega_A=|\vec{A}|^2.
\ea
On the $SU(N)$ side the corresponding basis is
\ba
T_A & = & N z^{\frac{1}{2} A_1 A_2}\Omega_2^{A_1}
\Omega_1^{A2} ,\non
\Omega_1\Omega_2 & = & z \Omega_2 \Omega_1,
\qquad z=e^{2\pi i/N}.
\ea
The analogue of the Laplacian in this theory may be
picked up by using the adjoint action of $\Omega$ as,
\ba
\widetilde{\Delta}  & \equiv & 
{}-\frac{N^2}{4\pi^2}\left(\mbox{Ad}{(\Omega_1)}+
\mbox{Ad}{(\Omega_2)}+
\mbox{Ad}{(\Omega_1^{-1})}+
\mbox{Ad}{(\Omega_2^{-1})}-4\right) ,\non
\widetilde{\Delta} T_{A} & = & -\widetilde\omega_A T_{A} ,\non
\widetilde{\omega}_A & = & -\frac{N^2}{4\pi^2}
(z^{A_1}+z^{A_2}+z^{-A_1}+z^{-A_2} -4).
\ea
In the large N limit, we restore the relation
in the continuous limit 
$\lim_{N\rightarrow \infty}\widetilde{\omega}_A=\omega_A$.
Obviously, in such a situation, one may define
the matrix model analogue of the Lorentz generator
in such a way that it gives the correct commutation relation.

The message here is that we need to specify
the Laplacian in the M(atrix) theory 
to define the Lorentz generators.
It should be encoded in the eleven dimensional definition
of $\cM$ theory and have to be tightly restricted
since otherwise the various identities discussed in 
appendix A will be violated and so is the Lorentz symmetry.
The situation reminds us of the fact
that the consistent string
background is depicted by the conformal invariance
which is closely related to the Lorentz symmetry
in the light cone gauge.
One might say that the background
dependence of M(atrix) theory appear here as
the choice of the Laplacian and the constraint on
it comes from the Lorentz invariance as
the tensor identities.

\vskip 5mm
\noindent{\bf Acknowledgements:} \hskip 3mm
We would like to thank
M. Ninomiya, M. Anazawa, A. Ishikawa, K. Sugiyama
for invaluable discussions, comments, and encouragements.
We are also obliged to S. Watamura for enjoyable
conversation on the non-commutative geometry.

\vskip 5mm
\noindent{\bf Note Added:} \hskip 3mm
After we submitted this paper, we are notified that
the Lorentz algebra was also computed in the
unpublished work \cite{r:Mel}.
We also added appendix F to clarify our argument in section 4.
We thank H. Nicolai for the information and the comments. 

\appendix
\section{APD Identities}

In this section we present several identities
satisfied by the APD tensors.
Let us recall the definitions of three
tensors in (\ref{eq:fabc}) and (\ref{eq:dabc}):
\begin{eqnarray}
f_{ABC}&=&\int d^{2}\sigma \sqrt{w(\sigma)}
          Y_{A}(\sigma)\{Y_{B}(\sigma),Y_{C}(\sigma)\} ,
 \label{eq:f}\\
d_{ABC}&=&\int d^{2}\sigma \sqrt{w(\sigma)} Y_{A}(\sigma)
                                 Y_{B}(\sigma) Y_{C}(\sigma) ,
 \label{eq:d}\\
c_{ABC}&=&-2\int d^{2}\sigma\sqrt{w(\sigma)}
          \frac{w^{rs}(\sigma)}{\omega_{A}}
          \partial _{r} Y_{A}(\sigma) Y_{B}(\sigma)
          \partial_{s} Y_{C}(\sigma) .
  \label{eq:c}
\end{eqnarray}
{}From the above definitions it is evident that $f_{ABC}$ is
totally antisymmetric and $d_{ABC}$ is totally
symmetric. 
The basis functions $Y_{A}$ satisfy the completeness relations,
\begin{eqnarray}
&&\sum_{A}Y^{A}(\sigma)Y_{A}(\rho)
   =\frac{1}{\sqrt{w(\sigma)}}\delta^{(2)}(\sigma ,\rho) -1 ,
    \label{eq:com1}\\
&&\sum_{A}\frac{1}{\omega_{A}}
     \left[ D^{r}Y_{A}(\sigma)D^{s}Y^{A}(\rho)
      +\frac{\epsilon^{rt}}{\sqrt{w(\sigma)}}\partial_{t}Y_{A}(\sigma)
           \frac{\epsilon^{su}}{\sqrt{w(\rho)}}\partial_{u}Y^{A}(\rho)
     \right] \nonumber \\
    &&\mbox{\hspace{15pc}}
    =\frac{w^{rs}(\sigma)}{\sqrt{w(\sigma)}}\delta^{(2)}(\sigma, \rho)
       {} -\sum_{\lambda}\Phi^{(\lambda)r}(\sigma)
                       \Phi^{(\lambda )s}(\rho).
    \label{eq:com2}
\end{eqnarray}
By using (\ref{eq:com1}) and integrating by parts,
the authors of \cite{r:dWMN} showed that the APD tensors satisfy
several identities:
\begin{eqnarray}
{f_{[AB}}^{E}f_{C]DE} &=& 0 ,\nonumber\\
c_{ABC}+c_{ACB} 
&=& 2\int d^{2}\sigma \sqrt{w(\sigma)}\frac{1}{\omega}_{A}
    \Delta Y_{A}(\sigma)Y_{B}(\sigma)Y_{C}(\sigma)
=-2d_{ABC},\nonumber\\
{c_{DE}}^{[A}f^{BC]E}
&=&2\int d^{2}\sigma \sqrt{w(\sigma)}
  \frac{w^{rs}(\sigma)}{\omega_{D}}\partial_{r}Y_{D}(\sigma)
  \partial_{s}Y^{[A}(\sigma)
  \{ Y^{C}(\sigma),Y^{B](\sigma)}\}
=0 ,\label{eq:wmn}\\
d_{ABC}{f^{A}}_{[DE}{f^{B}}_{F]G} 
&=&\int d^{2}\sigma \sqrt{w(\sigma)}
  Y_{C}(\sigma)\{Y_{[D}(\sigma),Y_{E}(\sigma)\}
  \{ Y_{F]}(\sigma),Y_{G}(\sigma)\}
=0,\nonumber\\
{f_{A(B}}^{E}d_{CD)E} 
&=&\displaystyle\int d^{2}\sigma \sqrt{w(\sigma)}
  \frac{1}{3}
  \{Y_{A}(\sigma)\,,\,Y_{B}(\sigma)Y_{C}(\sigma)Y_{D}(\sigma)\}
= 0 ,\nonumber\\
d_{EA[B}{d_{C]D}}^{E} 
&=&\displaystyle\int d^{2}\sigma \sqrt{w(\sigma)}
  Y_{A}Y_{[B}Y_{C]}Y_{D}
  {}-\displaystyle\int d^{2}\sigma \sqrt{w(\sigma)}
    Y_{A}Y_{[B}
  \displaystyle\int d^{2}\rho \sqrt{w(\rho)}
    Y_{C]}Y_{D}\nonumber\\
&=& -\eta_{A[B}\eta_{C]D}.\nonumber
\end{eqnarray}
The first identity is nothing but Jacobi identity.
In the last step to derive the third and fourth ones,
they used the fact that the parameter space
of the APD gauge group is two dimensional (Schouten's
identity).
The others are straightforward. They also found from
(\ref{eq:com2}) that
\begin{equation}
{f_{AB}}^{E}c_{ECD}
  = c_{EAB}{f^{E}}_{CD}-2{f_{BD}}^{E}d_{ACE}
    +\sum_{\lambda}c_{\lambda AB}{f^{\lambda}}_{CD},
  \label{eq:fc}
\end{equation}
where
\begin{equation}
\begin{array}{ccl}
f_{\lambda AB}&=& \displaystyle{\int} d^{2}\sigma \sqrt{w(\sigma)}
   \Phi^{(\lambda)t}(\sigma)\partial_{t}Y_{B}(\sigma)Y_{A}(\sigma) ,\\
c_{\lambda AB}&=&-2\displaystyle{\int} d^{2}\sigma \epsilon^{rs}
   \Phi_{r}^{(\lambda)}(\sigma)\partial_{s}Y_{B}(\sigma)Y_{A}(\sigma).
\end{array}
\end{equation}

Besides these identities we have derived
\begin{eqnarray}
&&-c^{ABC}{d_{C}}^{EF}+2c^{AC(E}{d_{C}}^{F)B}
   = 4\eta^{A(E}\eta^{F)B}, \label{eq:i} \\
&&\frac{1}{4}({c_{C}}^{FE}c^{[AB]C}+{c^{[A|F|}}_{C}c^{B]CE})
     \nonumber \\
&& \mbox{\hspace{3pc}}
      = -\frac{1}{2}\left(\frac{1}{\omega_{A}}-\frac{1}{\omega_{B}}
                    \right)
         \left(\frac{1}{\omega_{C}}{f_{C}}^{EF}f^{CAB}
          +{f_{\lambda}}^{EF}f^{\lambda AB}\right)
       +\frac{1}{2\omega_{A}\omega_{B}}{f_{C}}^{AB}f^{CEF},
     \label{eq:ii}\\
&&{d_{E}}^{AC}(c^{DEB}-2d^{DEB})-2c^{D(C|E|}{d_{E}}^{A)B}
    =4\eta^{AC}\eta^{BD}, \label{eq:iii} \\
&&{d^{CG}}_{H}{d^{DH}}_{I}f^{I[EF}{f^{B]A}}_{G}
   =-f^{D[EF}f^{B]AC}.
   \label{eq:iv}
\end{eqnarray}
They play very important roles
in our computation in the following sections.
Here we give their brief derivation.
Using (\ref{eq:com1}) and integrating by parts, we obtain
\begin{eqnarray}
c^{ABC}{d_{C}}^{EF}
&=&-2\int d^{2}\sigma\sqrt{w(\sigma)}
   \frac{D^{r}Y^{A}(\sigma)}{\omega_{A}}
   Y^{B}(\sigma)\partial_{r}(Y^{E}(\sigma)Y^{F}(\sigma)),\\
c^{ACE}{d_{C}}^{EF}
&=&2\eta^{AB}\eta^{BF}
   {}-2\int d^{2}\sigma\sqrt{w(\sigma)}
   \frac{D^{r}Y^{A}(\sigma)}{\omega_{A}}
   Y^{B}(\sigma)\partial_{r}Y^{E}(\sigma)Y^{F}(\sigma).
\end{eqnarray}
The combination of these relations gives (\ref{eq:i}).
In order to derive (\ref{eq:ii}) we rewrite the
first term in the l.h.s. of (\ref{eq:ii})  by
using (\ref{eq:com2}):
\begin{eqnarray}
\frac{1}{4}{c_{C}}^{FE}c^{ABC}
&=&\frac{1}{\omega_{A}}\int d^{2}\sigma\sqrt{w(\sigma)}
  D^{s}Y^{A}(\sigma)\partial_{s} Y^{E}(\sigma)Y^{F}(\sigma)
  Y^{B}(\sigma)\nonumber \\
  &&\mbox{\hspace{3pc}}
  {}-\frac{1}{\omega_{A}\omega_{C}}{f^{FE}}_{C}f^{BAC}
  {}-\frac{1}{\omega_{A}}{f_{\lambda}}^{EF}f^{\lambda AB}.
           \label{eq:ii1st}
\end{eqnarray}
The completeness relation (\ref{eq:com1}) and integration by parts
enable us to rewrite
the second term in the l.h.s. of (\ref{eq:ii}):
\begin{eqnarray}
\frac{1}{4}{c^{AF}}_{C}c^{BCE}
&=& \frac{1}{\omega_{A}}\int d^{2}\sigma\sqrt{w(\sigma)}
 D^{s}Y^{B}(\sigma)\partial_{s}Y^{E}(\sigma)Y^{F}(\sigma)Y^{A}(\sigma)
 \nonumber \\
 &&\mbox{\hspace{3pc}}
 {} -\frac{1}{\omega_{A}\omega_{B}}\int d^{2}\sigma\sqrt{w(\sigma)}
   D^{r}Y^{A}(\sigma)\partial_{r}Y_{F}(\sigma)
   D^{s}Y^{B}(\sigma)\partial_{s}Y^{E}(\sigma)\ .
    \label{eq:ii2nd}
\end{eqnarray}
{}From (\ref{eq:ii1st}) and (\ref{eq:ii2nd}) we obtain
\begin{eqnarray}
&&\frac{1}{4}({c_{C}}^{FE}c^{ABC}+{c^{AF}}_{C}c^{BCE})\nonumber\\
&&\mbox{\hspace{2pc}}=
{}-\frac{1}{\omega_{A}}
   \left( 
    \frac{1}{\omega_{C}}{f_{C}}^{EF}f^{CAB}
    +{f_{\lambda}}^{EF}f^{\lambda AB}
   \right) \nonumber\\
&&\mbox{\hspace{3pc}}
+\int d^{2}\sigma\sqrt{w(\sigma)}
   \partial_{s}Y^{E}(\sigma)Y^{F}(\sigma)
   \left(
     \frac{D^{s}Y^{A}(\sigma)}{\omega_{A}}Y^{B}(\sigma)
     +\frac{D^{s}Y^{B}(\sigma)}{\omega_{B}}Y^{A}(\sigma)
   \right)  \nonumber \\
&&\mbox{\hspace{3pc}}
{}-\frac{1}{\omega_{A}\omega_{B}}H^{ABFE},
\label{eq:ii3rd}
\end{eqnarray}
where $H^{ABFE}$ is defined by
\begin{equation}
H^{ABFE}=\int d^{2}\sigma\sqrt{w(\sigma)}
         D^{r}Y^{A}(\sigma)\partial_{r}Y^{F}(\sigma)
         D^{s}Y^{B}(\sigma)\partial_{s}Y^{E}(\sigma).
\end{equation}
When we antisymmetrize the indices $A$ and $B$ in (\ref{eq:ii3rd}),
the second term in the r.h.s of (\ref{eq:ii3rd}) vanishes
and the third term turns out to be
\begin{equation}
H^{[AB]FE}=-\frac{1}{2}{f_{C}}^{AB}f^{CEF}.
 \label{eq:ii4th}
\end{equation}
The derivation of  this equation requires the identity
\begin{equation}
\delta^{r}_{t}\delta^{s}_{u}-\delta^{s}_{t}\delta^{r}_{u}
  =\epsilon^{rs}\epsilon_{tu}.
\end{equation}
The combination of (\ref{eq:ii3rd}) and (\ref{eq:ii4th}) gives
(\ref{eq:ii}).
Next we derive (\ref{eq:iii}).
Eq.(\ref{eq:com1}) and integration by parts
give the following relation:
\begin{eqnarray}
&&{d_{E}}^{AC}(c^{DEB}-2d^{DEB})-2c^{DCE}{d_{E}}^{AB}\nonumber \\
&&\mbox{\hspace{2pc}}
  =4\eta^{AC}\eta^{BD}\nonumber\\
&&\mbox{\hspace{3pc}}
  {}-\int d^{2}\sigma \sqrt{w(\sigma)}\left[
   \frac{D^{r}Y_{D}}{\omega_{D}}\partial_{r}Y^{B}Y^{A}Y^{C}
    +Y^{D}Y^{B}Y^{A}Y^{C}
    {}-\frac{D^{r}Y_{D}}{\omega_{D}}Y^{C}\partial_{r}(Y^{A}Y^{B})
      \right] \nonumber \\
&&\mbox{\hspace{2pc}}
  =4\eta^{AC}\eta^{BD}-2\int d^{2}\sigma\sqrt{w(\sigma)}
  \frac{D^{r}Y^{D}}{\omega_{D}}Y^{B}
   (\partial_{r}Y^{C}Y^{A}
     {}-\partial_{r}Y^{A}Y^{C}).
\end{eqnarray}
{}From this relation we find the identity (\ref{eq:iii}).
{}Finally we prove the identity (\ref{eq:iv}) .
By using (\ref{eq:com1}) and integrating by parts we obtain
\begin{equation}
{d^{CG}}_{H}{d^{DH}}_{I}f^{IEF}f^{BAC}
=\int d^{2}\sigma\sqrt{w(\sigma)}Y^{C}Y^{D}
  \{Y^{E}\,,\,Y^{F}\}\{Y^{B}\,,Y^{A}\}
  {}-f^{DEF}f^{BAC}.
\end{equation}
The first term in the r.h.s. vanishes by antisymmetrizing
indices $B$, $E$ and $F$.

\section{Some Identities of $SO(9)$ Clifford Algebra}

In this section we review some properties of $SO(9)$ gamma
matrices ${\gamma^{a}}_{\alpha\beta}$ %
($a=1,\ldots ,9$ ; $\alpha , \beta =1,\ldots ,16$).
We can take ${\gamma^{a}}_{\alpha\beta}$
as real and symmetric matrices, i.e.
\begin{equation}
({\gamma^{a}}_{\alpha\beta})^{\ast}={\gamma^{a}}_{\alpha\beta}\ ,
\ \ {\gamma^{a}}_{\alpha\beta}={\gamma^{a}}_{\beta\alpha}.
\end{equation}
{}From these gamma matrices
we can construct an orthogonal complete basis of $16\times 16$
real matrices (or bilinear products of real spinors) :
\begin{equation}
\left\{ I_{\alpha\beta}\ ,
       \ {\gamma^{a}}_{\alpha\beta}\ ,
       \ {\gamma^{ab}}_{\alpha\beta}\ ,
       \ {\gamma^{abc}}_{\alpha\beta}\ ,
       \ {\gamma^{abcd}}_{\alpha\beta}\ \right\}, \label{eq:basis}
\end{equation}
where
\begin{equation}
\gamma^{a_{1}\cdots a_{k}}
   =\gamma^{[a_{1}}\gamma^{a_{2}}\cdots \gamma^{a_{k}]}.
\end{equation}
$I$, $\gamma^{a}$ and $\gamma^{abcd}$ are
symmetric, and $\gamma^{ab}$ and $\gamma^{abc}$ are antisymmetric
with respect to the spinorial indices.

The SO(9) gamma matrices satisfy several identities,
such as
\begin{equation}
\gamma^{a}\gamma^{b_{1}\cdots b_{k}}
  =\gamma^{ab_{1}\cdots b_{k}}
     +\sum_{l=1}^{k}(-)^{l-1}\delta^{ab_{l}}
             \gamma^{b_{1}\cdots\check{b}_{l}\cdots b_{k}}\ \ ,
  \label{eq:irr}
\end{equation}
\begin{eqnarray}
&&(\gamma^{b})_{\alpha\beta}(\gamma_{ab})_{\gamma\delta}
   +(\gamma^{b})_{\gamma\delta}(\gamma_{ab})_{\alpha\beta}
+(\gamma^{b})_{\alpha\delta}(\gamma_{ab})_{\gamma\beta}
+(\gamma^{b})_{\gamma\beta}(\gamma_{ab})_{\alpha\delta}
\nonumber \\
&&\mbox{\hspace{5pc}}-2I_{\delta\beta}(\gamma_{a})_{\gamma\alpha}
+2I_{\alpha\gamma}(\gamma_{a})_{\beta\delta}=0\ \ .
   \label{eq:fi}
\end{eqnarray}
By multiplying $(\gamma^{a})_{\delta\epsilon}
  (\theta_{[A}^{\beta}\theta_{B}^{\gamma}\theta_{C]}^{\epsilon})$ to
(\ref{eq:fi}) we derive
\begin{equation}
(\gamma_{d}\theta_{[A})^{\alpha}(\theta_{B}\gamma^{d}\theta_{C]})
  =\theta^{\alpha}_{[A}(\theta_{B}\theta_{C]})\ \ .
       \label{eq:fii}
\end{equation}
These identities of gamma matrices are useful in carrying 
out our calculation.


\section{$(\tilde{M}^{-a},\tilde{M}^{-b})_{DB}$}

In this section we show that (\ref{eq:MM}) holds
modulo the first class constraints $\varphi_{A}$ and
$\varphi_{\lambda}$.

{}First we substitute (\ref{eq:M2}) and (\ref{eq:Mt}) into
the first and second terms in the r.h.s. of (\ref{eq:C}).
The result is
\begin{equation}
{}-{\cal M}^{2}\tilde{M}^{ab}-\frac{i}{4}\tilde{Q}^{+}
\gamma^{ab}\tilde{Q}^{+}=C_{1}+C_{2}+C_{3}+C_{4}+C_{5}+C_{6},
\end{equation}
where each $C_{I}$ ($I=1,\ldots 6$) is composed of the
same type of polynomials  in $X$, $P$ and $\theta$:
\footnote{Note that we have expanded the products of
gamma matrices in the complete set (\ref{eq:basis}) by using the
identities such as (\ref{eq:irr}).}
\begin{eqnarray}
C_{1}&=&-\vec{P}_{A}^{2}(-P_{D}^{a}X^{bD}+P^{b}_{D}X^{aD}), \\
C_{2}&=&-\frac{i}{2}P^{a}_{A}P^{b}_{B}(\theta^{A}\theta^{B})
{}-\frac{i}{2}\{P^{a}_{A}(\theta^{A}\gamma^{bd}\theta^{B})
{}-P^{b}_{A}(\theta^{A}\gamma^{ad}\theta^{B})\}P_{dB}
\nonumber \\
&+&\frac{i}{4}\vec{P}^{2}_{A}(\theta_{B}\gamma^{ab}\theta^{B})
{}-\frac{i}{4}\vec{P}_{A}\cdot\vec{P}_{B}(\theta^{A}\gamma^{ab}\theta^{B})
{}-\frac{i}{4}{P}_{dA}{P}_{eB}(\theta^{A}\gamma^{abde}\theta^{B}), \\
C_{3}&=&-\frac{i}{2}f_{A}^{BC}X^{a}_{B}X^{b}_{C}
(\theta^{A}\gamma\cdot P_{D}\theta^{D})
\nonumber \\
&+&\frac{i}{2}f^{ABC}(X^{a}_{C}P^{b}_{D}-X^{b}_{C}P^{a}_{D})
(\theta_{A}\gamma\cdot X_{B}\theta^{D})
+if^{ABC}(X^{a}_{D}P^{bD}-X^{b}_{D}P^{aD})
(\theta_{A}\gamma\cdot X_{B}\theta_{C})
\nonumber \\
&-&\frac{i}{2}f^{ABG}\{X^{a}_{A}(\theta_{B}\gamma^{b}\theta^{H})
{}-X^{b}_{A}(\theta_{B}\gamma^{a}\theta^{H})\}\vec{X}_{G}\cdot\vec{P}_{H}
\nonumber \\
&-&\frac{i}{2}f^{ABC}\{X^{a}_{A}(\theta_{B}\gamma^{bde}\theta^{D})
{}-X^{b}_{A}(\theta_{B}\gamma^{ade}\theta^{D})\}X_{dC}P_{eD}
\nonumber \\
&-&\frac{i}{4}f^{ABC}\{P^{a}_{D}(\theta_{A}\gamma^{bde}\theta^{D})
{}-P^{b}_{D}(\theta_{A}\gamma^{ade}\theta^{D})\}X_{dB}X_{eC}
+\frac{i}{2}f^{ABG}(\theta_{A}\gamma^{abd}\theta^{H})
X_{dB}(\vec{X}_{G}\cdot\vec{P}_{H})
\nonumber \\
&+&\frac{i}{4}f^{ABC}(\theta_{A}\gamma^{abdef}\theta^{D})
X_{dB}X_{eC}P_{fD}, \\
C_{4}&=&\frac{1}{2}(f_{ABC}X^{B}_{d}X^{C}_{e})^{2}
(P^{a}_{D}X^{bD}-P^{b}_{D}X^{aD}), \\
C_{5}&=&\frac{i}{8}(\theta_{A}\gamma^{ab}\theta^{A})
(f_{BCD}X_{c}^{B}X_{d}^{D})^{2} 
\nonumber \\
&+&\frac{i}{16}(\theta^{B}\gamma^{cdabef}\theta^{C})
X_{c}^{D}X_{d}^{E}X_{e}^{F}X_{f}^{G}f_{BDE}f_{CFG}
\nonumber \\
&+&\frac{i}{16}\{\theta^{B}(-\delta^{ca}\gamma^{dbef}
+\delta^{cd}\gamma^{daef}-\delta^{ce}\gamma^{dabf}+
\delta^{cf}\gamma^{dabe}+\delta^{da}\gamma^{cbef}
{}-\delta^{db}\gamma^{caef}
\nonumber \\
& &+\delta^{de}\gamma^{cabf}-\delta^{df}\gamma^{cabe}
{}-\delta^{ae}\gamma^{cdbf}+\delta^{be}\gamma^{cdaf}+
\delta^{af}\gamma^{cdbe}-\delta^{bf}\gamma^{cdae})\theta^{C}\}
\nonumber \\
& &\qquad\quad \times
X^{D}_{c}X^{E}_{d}X^{F}_{e}X^{G}_{F}f_{BDE}f_{CFG}
\nonumber \\
&+&\frac{i}{8}(\theta^{B}\gamma^{cd}\theta^{C})
(X^{aE}X^{bD}-X^{aD}X^{bE})X^{F}_{c}X_{d}^{G}f_{BDE}f_{CFG}
\nonumber \\
&+&\frac{i}{4}(\theta^{B}\gamma^{cd}\theta^{C})
(X^{aF}X^{bD}-X^{bF}X^{aD})X_{c}^{E}X_{d}^{G}f_{BDE}f_{CFG}
\nonumber \\
&+&\frac{i}{8}(\theta^{B}\gamma^{ab}\theta^{C})
(\vec{X}^{D}\cdot\vec{X}^{G})(\vec{X}^{E}\cdot\vec{X}^{F})
f_{BDE}f_{CFG}
\nonumber \\
&+&\frac{i}{2}\{(\theta^{B}\gamma^{ae}\theta^{C})X^{bD}-
(\theta^{B}\gamma^{be}\theta^{C})X^{aD}\}X_{e}^{F}
(\vec{X}^{E}\cdot\vec{X}^{G})f_{BDE}f_{CFG}
\nonumber \\
&+&\frac{i}{4}(\theta^{B}\theta^{C})(X^{aE}X^{bF}-X^{aF}X^{bE})
(\vec{X}^{D}\cdot\vec{X}^{G})f_{BDE}f_{CFG}, \\
C_{6}&=&\frac{1}{4}f_{ABC}(\theta^{A}\gamma\cdot X^{B}\theta^{C})
(\theta_{D}\gamma^{ab}\theta^{D}).
\end{eqnarray}

Next we calculate $(\tilde{M}^{-a},\tilde{M}^{-b})_{DB}$.
By using the definition of the generator $\tilde{M}^{-a}$ in
(\ref{eq:Mt}) and the Dirac brackets (\ref{eq:DB}),
we find
\begin{equation}
(\tilde{M}^{-a},\tilde{M}^{-b})_{DB}=B_{1}+B_{2}+B_{3}+B_{4}+B_{5}
+B_{6},
\end{equation}
with $B_{1}$, $B_{2}$, ..., $B_{6}$ being given by:
\begin{eqnarray}
B_{1}&=&\frac{1}{4}(c^{DEB}d_{E}^{\;AC}-2c^{DCE}d_{E}^{\;AB}
{}-2d^{ACE}d_{E}^{\;BD})(X^{a}_{B}P^{b}_{D}-X^{b}_{B}P^{a}_{D})
\vec{P}_{A}\cdot\vec{P}_{C}\nonumber \\
&+&\frac{1}{4}(c^{AF}_{\quad C}c^{BCE}-c_{C}^{\;FE}c^{BAC})
\vec{X}_{E}\cdot\vec{P}_{F}(P^{a}_{A}P^{b}_{B}-P^{b}_{A}P^{a}_{B}),
\\
B_{2}&=&[-\frac{i}{4}d^{AEC}d_{C}^{\;BF}+\frac{i}{16}
\{2c_{C}^{\;EF}(c^{ABC}-c^{BAC})
\nonumber \\
& &\qquad+(c^{ACE}-c^{AEC})(c^{BFC}
{}-c^{BCF})\}]P^{a}_{A}P^{b}_{B}(\theta_{E}\theta_{F})
\nonumber \\
&+&[\frac{i}{8}\{-c^{ABC}d_{C}^{\;EF}+(c^{ACE}-c^{AEC})d_{C}^{\;BF}\}
{}-\frac{i}{4}d^{AEC}d_{C}^{BF}]\times
\nonumber \\
& &\qquad\qquad\times\{P^{a}_{A}(\theta_{E}\gamma^{bd}\theta_{F})
{}-P^{b}_{A}(\theta_{E}\gamma^{ad}\theta_{F})\}P_{dB}
\nonumber \\
&+&\frac{i}{4}(d^{ABC}d_{C}^{\;EF}-d^{AEC}d_{C}^{\;BF})
\vec{P}_{A}\cdot\vec{P}_{B}(\theta_{E}\gamma^{ab}\theta_{F})
\nonumber \\
&-&\frac{i}{4}d^{AEC}d_{C}^{\;BF}P_{dA}P_{eB}
(\theta_{E}\gamma^{abde}\theta_{F}), \\
B_{3}&=&-\frac{i}{4}(2d^{AD}_{\quad F}d^{BE}_{\quad G}f^{CFG}
+d^{AC}_{\quad G}d^{DB}_{\quad F}f^{EFG}-d^{AC}_{\quad G}
d^{DG}_{\quad F}f^{FBE})\times
\nonumber \\
& &\qquad\quad\times(\theta_{B}\gamma\cdot P_{A}\theta_{C})
(X^{a}_{D}X^{b}_{E}-X^{b}_{D}X^{a}_{E})\nonumber \\
&+&\frac{i}{8}\{2d^{C}_{\;FG}d^{EBF}f^{DAG}+4d^{DEF}d^{B}_{\;FG}f^{ACG}
{}-d^{DBF}f^{A}_{\;FG}(c^{ECG}-c^{EGC}) 
\nonumber \\
& &-2d^{DBF}f^{C}_{\;FG}c^{EGA}
{}-d_{G}^{\;DF}f_{F}^{\;AB}(c^{ECG}-c^{EGC})
{}-2d_{G}^{\;BF}f_{F}^{\;AC}c^{EGD}\}\times
\nonumber \\
& &\qquad\quad\times(\theta_{B}\gamma\cdot X_{A}\theta_{C})
(X^{a}_{D}P^{b}_{E}-X^{b}_{D}P^{a}_{E})
\nonumber \\
&+&\frac{i}{4}(d^{E}_{\;FG}d^{CDF}f^{ABG}+c^{FCA}d^{DBG}f^{E}_{\;FG}
+d^{BDF}d^{GCE}f^{A}_{\;FG}-d_{G}^{\;BF}d^{GCE}f_{F}^{\;DA})\times
\nonumber \\
& &\qquad\quad\times\{X^{a}_{B}(\theta_{D}\gamma^{b}\theta_{E})-
X^{b}_{B}(\theta_{D}\gamma^{a}\theta_{E})\}\vec{X}_{A}\cdot\vec{P}_{C}
\nonumber \\
&+&\frac{i}{4}(-d^{E}_{\;FG}d^{CDF}f^{ABG}+d^{A}_{\;FG}d^{CEG}f^{DBF}
{}-d_{F}^{\;AD}d_{G}^{\;CE}f^{FGB}+d_{F}^{\;AC}d_{G}^{\;DE}f^{BGF})\times
\nonumber \\
& &\times\{X^{a}_{A}(\theta_{D}\gamma^{bed}\theta_{E})
{}-X^{b}_{A}(\theta_{D}\gamma^{ade}\theta_{E})\}P_{dC}X_{eB}
\nonumber \\
&+&\frac{i}{16}\{(2d^{CDF}-c^{CFD}+c^{CDF})d^{E}_{\;FG}f^{ABG}
+2c^{CFA}d^{DEG}f^{B}_{\;FG}\}\times
\nonumber \\
& &\qquad\times\{P^{a}_{C}(\theta_{D}\gamma^{bed}\theta_{E})
{}-P^{b}_{C}(\theta_{D}\gamma^{aed}\theta_{E})\}X_{dA}X_{eB}
\nonumber \\
&+&\frac{i}{4}(c^{FED}d^{BCG}f^{A}_{\;FG}
{}-2d^{C}_{\;FG}d^{EBF}f^{DAG})(\theta_{B}\gamma^{abd}\theta_{C})
X_{dA}(\vec{X}_{D}\cdot\vec{P}_{E})
\nonumber \\
&-&\frac{i}{4}d^{E}_{\;FG}d^{CDF}f^{ABG}
(\theta_{D}\gamma^{abdef}\theta_{E})X_{dA}X_{eB}P_{fC}, \\
B_{4}&=&\frac{1}{8}(d_{H}^{\;IC}f_{I}^{\;DE}f_{C}^{\;FG}c^{BHA}+
4d^{AIC}f_{I}^{\;DE}f^{F}_{\;HC}c^{BHG}-2d^{ABC}d_{C}^{\;IH}f_{I}^{\;GD}
{}f_{H}^{\;EF})\times
\nonumber \\
& &\qquad\times(X^{a}_{A}P^{b}_{B}-X^{b}_{A}P^{a}_{B})(\vec{X}_{G}\cdot
\vec{X}_{E})(\vec{X}_{D}\cdot\vec{X}_{F})
\nonumber \\
&+&\frac{1}{2}(d^{AIC}f_{I}^{\;DE}f_{C}^{FH}c_{H}^{\;BG}-2d_{C}^{\;AB}
d^{EIH}f_{I}^{\;DG}f_{H}^{FC})\times \nonumber \\
& &\qquad\times(X^{a}_{A}X^{b}_{E}-X^{b}_{A}X^{a}_{E})
(\vec{X}_{D}\cdot\vec{X}_{F})(\vec{P}_{B}\cdot\vec{X}_{G}), \\
B_{5}&=&\frac{i}{16}(\theta_{A}\gamma^{acdbef}\theta_{B})
X_{cC}X_{dD}X_{eE}X_{fF}d^{A}_{\;GH}d^{BH}_{I}f^{CDG}f^{EFI}
\nonumber \\
&+&\frac{i}{16}\{\theta_{A}(-\delta^{ae}\gamma^{cdbf}
+\delta^{af}\gamma^{cdbe}-\delta^{cb}\gamma^{adef}+\delta^{ce}
\gamma^{adbf}
{}-\delta^{cf}\gamma^{adbe}+\delta^{db}\gamma^{acef}
\nonumber \\
& &-\delta^{de}\gamma^{acbf}+\delta^{df}\gamma^{acbe}
)\theta_{B}\}d^{A}_{\;GH}d_{I}^{\;BH}f^{CDG}f^{EFI}
\nonumber \\
&+&\frac{i}{8}\{(\theta_{D}\gamma^{ebcd}\theta_{F})X^{a}_{I}
{}-(\theta_{D}\gamma^{eacd}\theta_{F})X^{b}_{I}\}X_{cG}X_{dH}X_{eE}\times
\nonumber \\
& &\times(d_{C}^{\;DI}d^{F}_{\;AB}f^{AGH}f^{BCE}+f^{DEC}d^{I}_{\;CB}
d_{A}^{\;FB}f^{AGH})
\nonumber \\
&+&\frac{i}{8}(\theta_{A}\gamma^{ab}\theta_{B})(\vec{X}_{D}\cdot
\vec{X}_{E})(\vec{X}_{C}\cdot\vec{X}_{F})
(d^{A}_{\;GH}d_{I}^{\;BH}f^{CDG}f^{EFI}+
d_{G}^{\;AB}d^{G}_{\;HI}f^{ECH}f^{DFI})
\nonumber \\
&+&\frac{i}{4}\{(\theta_{A}\gamma^{ae}\theta_{B})X^{b}_{C}
{}-(\theta_{A}\gamma^{be}\theta_{B})X^{a}_{C}\}
(\vec{X}_{D}\cdot\vec{X}_{F})X_{eE}\times
\nonumber \\
& &\times(d^{A}_{\;GH}d_{I}^{\;BH}f^{CDG}f^{EFI}
{}-d_{G}^{\;AB}d^{C}_{\;HI}f^{DGH}f^{EFI}
\nonumber \\
& &\quad +d^{B}_{\;GH}d_{I}^{\;AC}f^{FEG}f^{DHI}
{}-d^{B}_{\;GH}d_{I}^{CH}f^{FEG}f^{DAI})
\nonumber \\
&+&\frac{i}{4}(\theta_{B}\gamma^{cd}\theta_{C})X^{a}_{F}X^{b}_{D}X^{c}_{E}
X^{d}_{G}\times
\nonumber \\
& &\times(f_{H}^{\;DE}d_{I}^{\;BH}d_{J}^{\;CI}f^{FGJ}
{}-f_{H}^{\;EB}d_{I}^{\;FH}d_{J}^{\;DI}f^{GCJ}
\nonumber \\
& &\quad+f_{H}^{\;EB}d_{I}^{\;FH}f_{J}^{\;GI}d^{CDJ}
{}-d_{H}^{\;BF}f_{I}^{\;EH}d_{J}^{\;DI}f^{GCJ})
\nonumber \\
& &\quad +d_{H}^{\;BF}f_{I}^{\;EH}f_{J}^{\;GI}d^{CDJ}
+f_{H}^{\;EG}d_{I}^{\;CH}f_{J}^{\;DI}d^{BFJ}
\nonumber \\
& &\quad-f_{H}^{\;EG}d_{I}^{\;CH}d_{J}^{\;FI}f^{DBJ})
\nonumber \\
&+&\frac{i}{8}(\theta_{A}\theta_{B})(X^{a}_{C}X^{b}_{D}
{}-X^{b}_{C}X^{a}_{D})(\vec{X}_{E}\cdot\vec{X}_{F})\times
\nonumber \\
& &\times(d^{A}_{\;GH}d_{I}^{\;BH}f^{FDG}f^{ECI}
{}-2c_{G}^{\;AB}d^{C}_{\;HI}f^{DEH}f^{FGI})
\nonumber \\
&+&\frac{i}{4}(\theta_{A}\theta_{D})(\vec{X}_{B}\cdot\vec{X}_{C})
(X^{a}_{E}X^{b}_{F})\times
\nonumber \\
& &\times(-f_{G}^{\;BA}d_{H}^{\;EG}d_{I}^{\;FH}f^{CDI}+
{}f_{G}^{\;BA}d_{H}^{\;EG}f_{I}^{\;CH}d^{DFI}
\nonumber \\
& &\quad-d_{H}^{\;AE}f_{G}^{\;BH}d_{I}^{\;FG}f^{CDI}
+d_{H}^{\;AE}f_{I}^{\;BH}f_{G}^{\;CI}d^{DFG}),  \\
B_{6}&=&\frac{1}{4}d^{ABC}f_{C}^{\;DE}d_{A}^{\;GH}
(\theta_{G}\gamma^{ab}\theta_{H})(\theta_{B}\gamma\cdot X_{D}\theta_{E})
\nonumber \\
&-&\frac{1}{8}d^{ABC}f_{C}^{\;DE}d_{D}^{\;GH}
\{X^{a}_{A}(\theta_{B}\gamma_{c}\theta_{E})(\theta_{G}\gamma^{bc}\theta_{H})
{}-X^{b}_{A}(\theta_{B}\gamma_{c}\theta_{E})(\theta_{G}\gamma^{ac}\theta_{H})\}
\nonumber \\
&+&\frac{1}{16}d^{ABC}f_{A}^{\;DE}d_{D}^{\;GH}X_{E}^{d}
\{(\theta_{B}\gamma^{ac}\theta_{C})(\theta_{G}\gamma^{bdc}\theta_{H})
{}-(\theta_{B}\gamma^{bc}\theta_{C})(\theta_{G}\gamma^{adc}\theta_{H})\}
\nonumber \\
&-&\frac{1}{8}d_{A}^{\;DE}f_{B}^{\;AC}c^{BGH}X_{cC}
(\theta_{D}\gamma^{abc}\theta_{E})(\theta_{G}\theta_{H})
\nonumber \\
&+&\frac{1}{8}d^{ABC}f_{C}^{\;DE}c_{D}^{\;GH}
\{X_{A}^{a}(\theta_{B}\gamma^{b}\theta_{E})-X^{b}_{A}
(\theta_{B}\gamma^{a}\theta_{E})\}(\theta_{G}\theta_{H}).
\end{eqnarray}

In order to see the cancellation between $C_{I}$ and $B_{I}$
($I=1,...,6$), it is convenient to rewrite $B_{I}$ by using
the identities in the appendices A and B.
Eq.(\ref{eq:iii}) reduces $B_{1}$ to
\begin{eqnarray}
B_{1}&=&\vec{P}_{A}^{2}(X^{a}_{B}P^{bB}-X^{b}_{B}P^{aB})
\nonumber \\
&+&\frac{1}{4}(c^{AF}_{\quad C}c^{BCE}+c_{C}^{\;FE}c^{ABC})
\vec{X}_{E}\cdot\vec{P}_{F}(P^{a}_{A}P^{b}_{B}-P^{b}_{A}P^{a}_{B}). 
\end{eqnarray}
By using (\ref{eq:wmn}),(\ref{eq:fc}) and (\ref{eq:i})
we rearrange $B_{2}$ and $B_{3}$ as
\begin{eqnarray}
B_{2}&=&-\frac{i}{4}[-2P^{a}_{A}P^{b}_{B}(\theta^{A}\theta^{B})
{}-2\{P^{a}_{A}(\theta^{A}\gamma^{bd}\theta^{B})-P^{b}_{A}
(\theta^{A}\gamma^{ad}\theta^{B})\}P_{dB}
\nonumber \\
& &+\vec{P}_{A}^{2}(\theta_{B}\gamma^{ab}\theta^{B})
{}-\vec{P}_{A}\cdot\vec{P}_{B}(\theta^{A}\gamma^{ab}\theta^{B})
{}-P_{dA}P_{eB}(\theta^{A}\gamma^{abde}\theta^{B})]
\nonumber \\
&+&\frac{i}{8}(c_{C}^{\;EF}c^{ABC}+c^{AE}_{\quad C}c^{BCF})
(P^{a}_{A}P^{b}_{B}-P^{b}_{A}P^{a}_{B})(\theta_{E}\theta_{F}) , \\
B_{3}&=&\frac{i}{2}f^{ABC}(\theta_{A}\gamma\cdot P_{D}\theta^{D})
(X^{a}_{B}X^{b}_{C})
\nonumber \\
&-&\frac{i}{2}f^{ABC}(\theta_{A}\gamma\cdot X_{B}\theta^{D})
(X^{a}_{C}P^{b}_{D}-X^{b}_{C}P^{a}_{D})
{}-if^{ABC}(\theta_{A}\gamma\cdot X_{B}\theta_{C})
(X^{a}_{D}P^{bD}-X^{b}_{D}P^{aD})
\nonumber \\
&+&\frac{i}{2}f^{ABG}\{X^{a}_{A}(\theta_{B}\gamma^{b}\theta^{H})
{}-X^{b}_{A}(\theta_{B}\gamma^{a}\theta^{H})\}\vec{X}_{G}\cdot\vec{P}_{H}
\nonumber \\
& & +\frac{i}{2}f^{DGH}\{X^{a}_{A}(\theta^{A}\gamma^{b}\theta_{D})
{}-X^{b}_{A}(\theta^{A}\gamma^{a}\theta_{D})\}\vec{X}_{G}\cdot\vec{P}_{H}
\nonumber \\
& &-\frac{i}{4}d^{ABC}(c_{DC}^{\quad E}f^{DGH}
+c_{\lambda C}^{\quad E}f^{\lambda GH})\{X^{a}_{A}
(\theta_{B}\gamma^{b}\theta_{E})-X^{b}_{A}
(\theta_{B}\gamma^{a}\theta_{E})\}\vec{X}_{G}\cdot\vec{P}_{H}
\nonumber \\
&+&\frac{i}{2}f^{ABC}\{X^{a}_{A}(\theta_{B}\gamma^{bde}\theta^{D})
{}-X^{b}_{A}(\theta_{B}\gamma^{ade}\theta^{D})\}X_{dC}P_{eD}
\nonumber \\
&+&\frac{i}{4}f^{ABC}\{P^{a}_{D}(\theta_{A}\gamma^{bde}\theta^{D})
{}-P^{b}_{D}(\theta_{A}\gamma^{ade}\theta^{D})\}X_{dB}X_{eC}
\nonumber \\
&-&\frac{i}{2}f^{ABG}(\theta_{A}\gamma^{abd}\theta^{H})X_{dB}
(\vec{X}_{G}\cdot\vec{P}_{H})+\frac{i}{2}f^{EGH}
(\theta_{D}\gamma^{abd}\theta^{D})X_{dE}(\vec{X}_{G}\cdot\vec{P}_{H})
\nonumber \\
& &-\frac{i}{4}d_{A}^{\;DE}(c_{B}^{\;AC}f^{BGH}+c_{\lambda}^{\;AC}
f^{\lambda GH})(\theta_{D}\gamma^{abd}\theta_{E})X_{dC}
(\vec{X}_{G}\cdot\vec{P}_{H})
\nonumber \\
&-&\frac{i}{4}f^{ABC}(\theta_{A}\gamma^{abdef}\theta_{D})
(X_{dB}X_{eC}P_{f}^{D}).
\end{eqnarray}
$B_{4}$ is rewritten by using (\ref{eq:wmn}), (\ref{eq:fc})
and (\ref{eq:iii}):
\begin{eqnarray}
B_{4}&=&\frac{1}{2}(f_{ABC}X_{d}^{B}X_{e}^{C})^{2}
(X^{a}_{D}P^{bD}-X^{b}_{D}P^{aD})
\nonumber \\
&+&\frac{1}{2}d^{AIC}f_{I}^{\;DE}(X^{a}_{A}X^{b}_{E}-X^{b}_{A}X^{a}_{E})
\vec{X}_{D}\cdot\vec{X}_{F}(c_{HC}^{\quad F}f^{HBG}+
c_{\lambda C}^{\quad F}f^{\lambda BG})\vec{P}_{B}\cdot\vec{X}_{G}
\nonumber \\
& &\quad +f^{DAE}(X^{a}_{A}X^{b}_{E}-X^{b}_{A}X^{a}_{E})
\vec{X}_{D}\cdot\vec{X}_{F}f^{FBG}\vec{P}_{B}\cdot\vec{X}_{G}.
\end{eqnarray}
We adopt (\ref{eq:wmn}), (\ref{eq:fc}) and (\ref{eq:iv})
to rewrite $B_{5}$ as
\begin{eqnarray}
B_{5}&=&-\frac{i}{16}(\theta_{B}\gamma^{abcdef}\theta_{C})
(X_{cD}X_{dE}X_{eF}X_{fG})f^{BDE}f^{CFG}
\nonumber \\
&-&\frac{i}{16}\{\theta_{B}(-\delta^{ae}\gamma^{cdbf}
+\delta^{af}\gamma^{cdbe}-\delta^{cb}\gamma^{adef}
+\delta^{ce}\gamma^{adbf}-\delta^{cf}\gamma^{adbe}
\nonumber \\
& &+\delta^{db}\gamma^{acef}-\delta^{de}\gamma^{acbf}
+\delta^{df}\gamma^{acbe})\theta_{C}\}
X_{cD}X_{dE}X_{eF}X_{fG}f^{BDE}f^{CFG}
\nonumber \\
&-&\frac{i}{8}\{(\theta_{D}\gamma^{ebcd}\theta_{F})X^{a}_{I}
{}-(\theta_{D}\gamma^{eacd}\theta_{F})X^{b}_{I}\}X_{cG}X_{dH}X_{eE}
{}f^{DEI}f^{FGH}
\nonumber \\
&-&\frac{i}{8}(\theta_{A}\gamma^{ab}\theta_{B})
(\vec{X}_{D}\cdot\vec{X}_{G})(\vec{X}_{E}\cdot\vec{X}_{F})
{}f^{EDA}f^{GFB}
{}-\frac{i}{8}(\theta_{A}\gamma^{ab}\theta^{A})
(f_{I}^{\;CD}X_{cC}X_{dD})^{2}
\nonumber \\
&-&\frac{i}{2}\{(\theta_{A}\gamma^{ae}\theta_{B})X^{b}_{C}
{}-(\theta_{A}\gamma^{be}\theta_{B})X^{a}_{C}\}
(\vec{X}_{D}\cdot\vec{X}_{F})X_{eE}f^{CDA}f^{EFB}
\nonumber \\
&-&\frac{i}{4}(\theta_{B}\gamma^{cd}\theta_{C})
(X^{a}_{F}X^{b}_{D}-X^{a}_{D}X^{b}_{F})X_{cE}X_{dG}f^{BDE}f^{CFG}
\nonumber \\
& &+\frac{i}{8}(\theta_{B}\gamma^{cd}\theta_{C})
(X^{a}_{D}X^{b}_{E}-X^{b}_{D}X^{a}_{E})(X_{cF}X_{dG})
{}f^{BDE}f^{CFG}
\nonumber \\
&+&\frac{i}{2}(\theta_{B}\theta_{C})(\vec{X}_{E}\cdot\vec{X}_{F})
X^{a}_{D}X^{b}_{G}(f^{BDE}f^{CGF}+2f^{EDG}f^{FBC})
\nonumber \\
& &+\frac{i}{4}d^{AKC}f_{K}^{\;DE}
(X^{a}_{A}X^{b}_{E}-X^{b}_{A}X^{a}_{E})(\vec{X}_{D}\cdot\vec{X}_{F})
(c_{HC}^{\quad F}f^{HBG}+c_{\lambda C}^{\quad F}f^{\lambda BG})
(\theta_{B}\theta_{G}).
\end{eqnarray}
{}Finally, rearrangement of $B_{6}$ requires (\ref{eq:wmn}),
(\ref{eq:fc}) and (\ref{eq:fi}). The result is
\begin{eqnarray}
B_{6}&=&-\frac{1}{4}f_{ABC}(\theta^{A}\gamma\cdot X^{B}\theta^{C})
(\theta_{D}\gamma^{ab}\theta^{D})
\nonumber \\
&-&\frac{1}{8}d_{A}^{\;DE}(\theta_{D}\gamma^{abd}\theta_{E})X_{dC}
\{c_{B}^{\;AC}f^{BGH}(\theta_{G}\theta_{H})+c_{\lambda}^{\;AC}
{}f^{\lambda GH}(\theta_{G}\theta_{H})\}
\nonumber \\
&+&\frac{1}{4}(\theta_{D}\gamma^{abd}\theta^{D})X_{dE}f^{EGH}
(\theta_{G}\theta_{H})
\nonumber \\
&-&\frac{1}{8}d^{ABC}\{X^{a}_{A}(\theta_{B}\gamma^{b}\theta_{E})-
X^{b}_{A}(\theta_{B}\gamma^{a}\theta_{E})\}
\{c_{DC}^{\quad E}f^{DGH}(\theta_{G}\theta_{H})+c_{\lambda C}^{\quad E}
f^{\lambda GH}(\theta_{G}\theta_{H})\}
\nonumber \\
&+&\frac{1}{4}\{X^{a}_{A}(\theta^{A}\gamma^{b}\theta_{D})-
X^{b}_{A}(\theta^{A}\gamma^{a}\theta_{D})\}f^{DGH}(\theta_{G}\theta_{H}).
\end{eqnarray}

We are now in a position to verify the cancellation
of (\ref{eq:C})  modulo the first class
constraints $(\varphi_{A},\varphi_{\lambda})$.
{}For this purpose we separate (\ref{eq:C}) into
three parts, namely,
\begin{eqnarray}
C&=&C^{(1)}+C^{(2)}+C^{(3)},\\
C^{(1)}&\equiv&B_{3}+C_{3}+B_{6}+C_{6},\nonumber \\
C^{(2)}&\equiv&B_{1}+C_{1}+B_{2}+C_{2},\nonumber \\
C^{(3)}&\equiv&B_{4}+C_{4}+B_{5}+C_{5}.\nonumber
\end{eqnarray}
After simple calculation we find
\begin{eqnarray}
C^{(1)}&=&\frac{i}{2}\{X^{a}_{A}(\theta^{A}\gamma^{b}\theta_{D})
{}-X^{b}_{A}(\theta^{A}\gamma^{a}\theta_{D})\}\varphi^{D}
\nonumber \\
& &-\frac{i}{4}\{X^{a}_{A}(\theta_{B}\gamma^{b}\theta_{E})
{}-X^{b}_{A}(\theta_{B}\gamma^{a}\theta^{E})\}
d^{ABC}(c_{DC}^{\quad E}\varphi^{D}+c_{\lambda C}^{\quad E}
\varphi^{\lambda})
\nonumber \\
&+&\frac{i}{2}(\theta_{D}\gamma^{abd}\theta^{D})X_{dE}\varphi^{E}
{}-\frac{i}{4}(\theta_{D}\gamma^{abd}\theta_{E})X_{dC}d_{A}^{\;DE}
(c_{B}^{\;AC}\varphi^{B}+c_{\lambda}^{\;AC}\varphi^{\lambda}), \\
C^{(2)}&=&\frac{1}{4}(c^{AF}_{\quad C}c^{BCE}+c_{C}^{\;FE}c^{ABC})
(\vec{X}_{E}\cdot\vec{P}_{F}-\frac{i}{2}\theta_{E}\theta_{F})
(P^{a}_{A}P^{b}_{B}-P^{b}_{A}P^{a}_{B}) ,\\
C^{(3)}&=&-\frac{1}{2}d^{AIC}f_{I}^{DE}(X^{a}_{A}X^{b}_{E}
{}-X^{b}_{A}X^{a}_{E})\vec{X}_{D}\cdot\vec{X}_{F}(c_{HC}^{\quad F}\varphi^{H}
+c_{\lambda C}^{\quad F}\varphi^{\lambda})
\nonumber \\
& &\qquad-f^{DAE}(X^{a}_{A}X^{b}_{E}-X^{b}_{A}X^{a}_{E})
\vec{X}_{D}\cdot\vec{X}_{F}\varphi^{F}.
\end{eqnarray}
We see that $C^{(1)}$ and $C^{(3)}$ are already written in the form of
linear combinations of $\varphi_{A}$ and $\varphi_{\lambda}$.
In order to show that $C^{(2)}$ is also linear in
the constraints, we have to apply
(\ref{eq:ii}). The result is
\begin{equation}
C^{(2)}=\{-\frac{1}{\omega_{A}}(\frac{1}{\omega_{C}}f^{CAB}\varphi_{C}
+f^{\lambda AB}\varphi_{\lambda})+\frac{1}{2\omega_{A}\omega_{B}}
f_{C}^{\;AB}\varphi^{C}\}(P_{A}^{a}P^{b}_{B}-P^{b}_{A}P^{a}_{B}).
\end{equation}

Thus we have proved that all the terms in (\ref{eq:C})
sum up to give (\ref{eq:MtMt}).

\section{$(\tilde{Q}^{+},\tilde{M}^{-a})_{DB}$}

In this section we show that the relation %
$(\tilde{Q}^{+},\tilde{M}^{-a})_{DB} = 0$ %
holds modulo the first class constraints.
By using relations (\ref{eq:DB}),
we can write down the result in the following way,
\begin{equation}
(\tilde{Q}^{+},\tilde{M}^{-a})_{DB}
     =D_{1}+D_{2}+D_{3}+D_{4},
\end{equation}
where
\begin{eqnarray}
(D_{1})_{\alpha} &=& -\frac{1}{2}(\gamma^{a}\theta_{A})_{\alpha}
             \vec{P}_{B}\cdot\vec{P}_{C}d^{ABC}
       {} -\frac{1}{4}(\gamma\cdot P_{A}\theta_{B})_{\alpha}P^{a}_{C}
              (c^{CBA}+c^{CAB})\nonumber\\
        &&\ \ \ 
              +\frac{1}{2}(\gamma\cdot P_{A}\gamma^{da}\theta_{C})_{\alpha}
              P_{dB}d^{ABC},\\
(D_{2})_{\alpha} &=& -\frac{1}{4}(\gamma^{a}\theta_{A})_{\alpha}
            (\vec X_{B}\cdot\vec X_{C})(\vec X_{D}\cdot\vec X_{E})
            {d^{A}}_{FG}f^{BDF}f^{CEG} \nonumber \\
        && +\frac{1}{8}(\gamma\cdot X_{A}\gamma\cdot X_{B}
                       \gamma^{ade}\theta_{C})_{\alpha}
               X_{dD}X_{eE}{d^{C}}_{FG}f^{ABG}f^{EDF} \nonumber\\
        && -\frac{1}{2}(\gamma\cdot X_{A}\theta_{B})_{\alpha}X^{a}_{C}
               (\vec X_{D}\cdot \vec X_{E})
               ({d^{C}}_{FG}f^{ADF}f^{BEG}+{d^{C}}_{FG}f^{BDF}f^{AEG})
               \nonumber\\
        &&-\frac{1}{4}(\gamma\cdot X_{A}\gamma\cdot X_{B}
                       \gamma\cdot X_{C} \theta_{D})_{\alpha}X^{a}_{E}
              ({d^{E}}_{FG}f^{ABF}f^{CDG}+{d^{DE}}_{F}{f^{AB}}_{G}f^{CFG}),\\
(D_{3})_{\alpha} &=& \frac{1}{2}
          (\gamma\cdot P_{A}\gamma\cdot X_{B}\theta_{C})_{\alpha}
          X^{a}_{D}({d^{CD}}_{F}f^{BAF}) \nonumber \\
        &&-\frac{1}{2}(\gamma\cdot X_{B}\gamma\cdot P_{A}
                            \theta_{C})_{\alpha}
            X^{a}_{D}{d^{AD}}_{F}f^{BCF} \nonumber \\
        && +\frac{1}{2}(\gamma^{ab}\theta_{A})_{\alpha}
          X_{dB}(\vec X_{C}\cdot\vec P_{D})c^{EDC}{f^{BA}}_{E}
             \nonumber\\
        &&-\frac{1}{4}
         \left[(\gamma\cdot P_{A}\gamma^{ade}\theta_{B})_{\alpha}
                   X_{dC}X_{eD}
           +(\gamma\cdot X_{C}\gamma\cdot X_{D}\gamma^{ab}\theta_{B}%
             )_{\alpha}P_{dA}\right]{d^{AB}}_{E}f^{CDE} \nonumber\\
        &&+\frac{1}{8}(\gamma\cdot X_{A}\gamma\cdot X_{B}\theta_{C}%
             )_{\alpha}P^{a}_{D}\nonumber\\
        &&\ \ \ \ \ \ \ \times(-2c^{DEB}{f^{AC}}_{E}+2c^{DEA}{f^{BC}}_{E}
                   +c^{DEC}{f^{AB}}_{E}-c^{DCE}{f^{AB}}_{E}),\\
(D_{4})_{\alpha} &=&  \frac{i}{2}(\gamma^{a}\theta_{A})_{\alpha}
              (\theta_{B}\gamma\cdot
                  X_{C}\theta_{D}){d^{AB}}_{E}f^{CDE}\nonumber\\ 
        && +\frac{i}{4}(\gamma^{ad}\theta_{B})_{\alpha}X_{dA}
              (\theta_{C}\theta_{D})c^{ECD}{f^{AB}}_{E}\nonumber\\
        && +\frac{i}{4}\left[-(\gamma_{d}\theta_{A})_{\alpha}
                         (\theta_{B}\gamma^{ade}\theta_{C})
                       +(\gamma^{ed}\theta_{A})_{\alpha}
                         (\theta_{B}{\gamma^{a}}_{d}\theta_{C})\right]
                   X_{eD}{d^{BC}}_{E}f^{DAE} \nonumber\\
        && +\frac{i}{2}(\gamma_{d}\theta_{A})_{\alpha}
               (\theta_{B}\gamma^{d}\theta_{C})X^{a}_{D}
                 {d^{BD}}_{E}f^{ACE}.\label{eq:D4}
\end{eqnarray}

After decomposing the products of gamma
matrices 
into the complete basis (\ref{eq:basis}),
the identities (\ref{eq:wmn}) and (\ref{eq:fc}) lead
us to find that $D_{1}$ and $D_{2}$ vanish and
\begin{equation}
(D_{3})_{\alpha}=
      \theta_{\alpha C}(\vec X_{B}\cdot\vec P_{A})X^{a}_{D}
          {d^{CD}}_{F}f^{BAF}
       +\frac{1}{2}(\gamma^{ab}\theta_{A})_{\alpha}X_{dB}
        (\vec X_{C}\cdot\vec P_{D})
        ({c_{E}}^{AB}f^{ECD}+{c_{\lambda}}^{AB}f^{\lambda CD}).
   \label{eq:D3}
\end{equation}
The third term in the r.h.s of (\ref{eq:D4}) is rewritten as
\begin{equation}
\frac{i}{4}\left[(\gamma_{d})_{\alpha\beta}
                  (\gamma^{ed})_{\gamma\delta}
            +(\gamma^{ed})_{\alpha\beta}(\gamma_{d})_{\gamma\delta}
            \right]\theta_{\beta A}(\theta_{B}\gamma^{a})_{\gamma}
            \theta_{\delta C}X_{eD}{d^{BC}}_{E}f^{DAE}.
\end{equation}
The identity (\ref{eq:fi}) and similar calculations enable us to
rewrite this term as
\begin{eqnarray}
&&\frac{i}{6}\left[
   (\gamma_{d}\theta_{C})_{\alpha}(\theta_{B}\gamma^{d}\theta_{A})
    +2(\theta_{C}\theta_{A})\theta_{\alpha B} \right]
     X^{a}_{D}{d^{BC}}_{E}f^{DAE}  \nonumber \\
&&\ \ \ +\frac{i}{2}\left[
       (\gamma^{a}\theta_{C})_{\alpha}(\theta_{A}\gamma^{e}\theta_{B})
      +(\theta_{B}\theta_{A})(\gamma^{ea}\theta_{C})_{\alpha}\right]
        X_{eD}{d^{BC}}_{E}f^{DAE}.
\end{eqnarray}
After combining with this term the other two terms in (\ref{eq:D4})
and using (\ref{eq:fc}),
\begin{eqnarray}
(D_{4})_{\alpha} &=&
  {}-\frac{i}{4}(\gamma^{ad}\theta_{C})_{\alpha}X_{dD}
    (\theta_{B}\theta_{A})
    ({c_{E}}^{CD}f^{EBA}+{c_{\lambda}}^{CD}f^{\lambda BA})
     \nonumber\\
    &&-\frac{i}{3}\left[(\theta_{B}\theta_{A})\theta_{\alpha C}
            +(\theta_{C}\theta_{A})\theta_{\alpha B} \right]
        X^{a}_{D}{d^{BD}}_{E}f^{CAE} \nonumber \\
    && -\frac{i}{2}\left[(\gamma_{d}\theta_{[A})^{\alpha}
            (\theta_{B}\gamma^{d}\theta_{C]})\right]
            X^{a}_{D}{d^{BD}}_{E}f^{CAE}.
\end{eqnarray}
By using (\ref{eq:fii}) in the third term of the r.h.s. of the
above equation, we finally obtain
\begin{equation}
(D_{4})_{\alpha} =
  {}-\frac{i}{4}(\gamma^{ad}\theta_{C})_{\alpha}X_{dD}
    (\theta_{B}\theta_{A})({c_{E}}^{CD}f^{EBA}
            +{c_{\lambda}}^{CD}f^{\lambda BA}) \nonumber\\
    \ \ -\frac{i}{2}(\theta_{C}\theta_{A})
       \theta_{\alpha B}{d^{BD}}_{E}f^{CAE}.\label{eq:D42}
\end{equation}
{}From (\ref{eq:D3}) and (\ref{eq:D42}) we conclude that
\begin{equation}
(\tilde{Q}^{+}_{\alpha},\tilde{M}^{-a})_{DB}
=\theta_{\alpha C}X^{a}_{D}{d^{BD}}_{E}\varphi^{E}
  +\frac{1}{2}(\gamma^{ad}\theta_{C})_{\alpha}X_{dD}
    ({c_{E}}^{CD}\varphi^{E}+{c_{\lambda}}^{CD}\varphi^{\lambda}).
\end{equation}

\section{$c_{ABC}$ in terms of $d_{ABC}$}

In this section we derive the equation (\ref{eq:cd})
which describes $c_{ABC}$ in terms of
the invariant tensor $d_{ABC}$.

Let us recall one of the relations in (\ref{eq:wmn})
\begin{equation}
c_{ABC}+c_{ACB}=-2d_{ABC}.
\end{equation}
By using the definitions of $c_{ABC}$ and $d_{ABC}$
and performing integration by parts, we find
another relation:
\begin{eqnarray}
c_{ABC}-c_{ACB}&=&2\int d^{2}\sigma\frac{1}{\omega_{A}}
Y_{A}(Y_{B}\Delta Y_{C}-Y_{C}\Delta Y_{B})
\nonumber \\
&=&2\frac{\omega_{B}-\omega_{C}}{\omega_{A}}d_{ABC}. 
\end{eqnarray}
Combining these two relations we can express
$c_{ABC}$ in terms of $d_{ABC}$:
\begin{equation}
c_{ABC}=(\frac{\omega_{B}-\omega_{C}}{\omega_{A}}-1)d_{ABC}. 
\end{equation}
This completes the proof of (\ref{eq:cd}).

\section{Supersymmetry algebra in M(atrix) theory}

The authors of \cite{r:BSS} observed some discrepancies 
in the supercharge algebra
between dWHN model and M(atrix) theory.
In this section, we would like to indicate that
there are some missing terms in the M(atrix) computation
and the alleged discrepancy can be removed.

We define
the supercharge $Q$ and the Dirac brackets between
the canonical variables $X$, $P$, $\theta$ of M(atrix) theory
in the same way as \cite{r:BSS}.
By using their defining relations, we obtain
\begin{eqnarray}
\{Q_{\alpha}\ , \ Q_{\beta}\}_{DB}&=&4RH\delta_{\alpha\beta}
 \nonumber\\
&&+2R\mbox{Tr}\left( -i
  \{P^{a}\, ,\, [X_{a}\, ,\, X_{b}]\}-i
  \left[\, [X_{b}\, , \, \theta^{\alpha^{\prime}}]
  \, ,\,
  \theta_{\alpha^{\prime}}\right]\right)(\gamma^{b})_{\alpha\beta}
 \nonumber\\
&&+2R\mbox{Tr}\left( X^{[a}X^{b}X^{c}X^{d]}\right)
   (\gamma_{abcd})_{\alpha\beta}\nonumber\\
&&-4iR\mbox{Tr}\left( [X_{b}\, , \, \frac{3}{8}
   \theta\gamma^{ab}\theta]\right)
   (\gamma_{a})_{\alpha\beta}\nonumber\\
&&+4iR\mbox{Tr}\left( [ X_{[a}\, ,\, \frac{1}{48}
  (\theta\gamma_{bcd]}\theta)]\right)
  (\gamma^{abcd})_{\alpha\beta}.
\end{eqnarray}
This result coincides with that of dWHN \cite{r:dWHN}
except that $\mbox{Tr}\left( X^{[a}X^{b}X^{c}X^{d]}\right)$
term is automatically vanishing in
the supermembrane calculation.
The last two $\theta$-bilinear terms in the r.h.s. of  the above
equation were absent in \cite{r:BSS}.
These two terms are originated from the second term
in the r.h.s. of the following relation,
\begin{equation}
[{{P^{a}}_{i}}^{j}\ ,\ Q_{\alpha}]_{DB}
=-i\sqrt{R}\left(2{[X_{d}\, ,\, \theta^{\alpha^{\prime}}]_{i}}^{j}
  (\gamma^{ab})_{\alpha\alpha^{\prime}}+
  \mbox{Tr}[\{D({{}^{j}}_{i})\, ,\, \theta^{\alpha^{\prime}}\}
  \, ,\, X_{d}](\gamma^{ad})_{\alpha\alpha^{\prime}}\right),
\label{eq:PQ}
\end{equation}
where $D({{}^{j}}_{i})$ is matrix valued quantity whose
$(k,p)$-component is defined as
\begin{equation}
{D({{}^{j}}_{i})_{k}}^{p}={\delta_{k}}^{j}{\delta_{i}}^{p}.
\end{equation}
By keeping this term, we have recovered the $\theta$-bilinear terms.

Here we should make a remark. The recoverd $\theta$-bilinear
terms cannot be observed  even if
we consider topologically nontrivial configurations
such as winding sectors of $\{X^{a}\}$.
This is because these terms are originated from the second term of
(\ref{eq:PQ}) which is expected to vanish in the configurations
with a well-defined supercharge $Q_{\alpha}$.
In this respect, it is proper to say that
there is no discrepancy between the result of dWHN \cite{r:dWHN}
and that of \cite{r:BSS}, at least practically.



\end{document}